\documentclass{aa}

\usepackage{txfonts}
\usepackage{graphicx}
\usepackage{astrobib}

\begin{document}

  \title{On the nature of late X-ray flares in \emph{Swift} Gamma-ray Bursts}
  
  \author{
    P.A.~Curran\inst{1} 
    \and R.L.C.~Starling\inst{2}
    \and P.T.~O'Brien\inst{2}
    \and O.~Godet\inst{2}
    \and A.J.~van~der~Horst\inst{3} 
    \and R.A.M.J.~Wijers\inst{1}
  }

  \offprints{P.A.~Curran \email{pcurran@science.uva.nl}}

\institute{
  Astronomical~Institute, University~of~Amsterdam, Kruislaan~403, 1098~SJ~Amsterdam, The~Netherlands
  \and Department~of~Physics~and~Astronomy, University~of~Leicester, University~Road, Leicester~LE1~7RH, UK
  \and NASA~Postdoctoral~Program~Fellow, NSSTC, 320~Sparkman~Drive, Huntsville, AL~35805, USA
}

\date{Received / Accepted }


\abstract
{Previously detected in only a few gamma-ray bursts (GRBs), X-ray flares are now observed in $\sim 50$\% of \emph{Swift} GRBs, though their origins remain unclear. Most flares are seen early on in the afterglow decay, while some bursts exhibit flares at late times of $10^4$ to $10^5$ seconds, which may have implications for flare models.}
{We investigate whether a sample of late time ($\gtrsim 1 \times 10^4$\,s) flares are different from previous samples of early time flares, or whether they are merely examples on the tail of the early flare distribution.}
{We examine the X-ray light curves of \emph{Swift} bursts for late flares and compare the flare and underlying temporal power-law properties with those of early flares, and the values of these properties predicted by the blast wave model. }
{The burst sample shows late flare properties consistent with those of early flares, where the majority of the flares can be explained by either internal or external shock, though in a few cases one origin is favoured over the other. The underlying power laws are mainly consistent with the normal decay phases of the afterglow. }
{If confirmed by the ever growing sample of late time flares, this would imply that, in some cases, prolonged activity out to a day or a restarting of the central engine is required.}


\maketitle

\titlerunning{On the nature of late X-ray flares in \emph{Swift} GRBs}
\authorrunning{P.A.~Curran et al.}


\section{Introduction}\label{section:introduction}

The majority of gamma-ray bursts (GRBs) are well described by the blast wave model \cite{rees1992:mnras258,meszaros1998:apj499}, which details their temporal and spectral behaviour. In this model GRB prompt emission is caused by internal shocks within a collimated ultrarelativistic jet while afterglow emission is created by external shocks when the jet ploughs into the circumburst medium, causing a blast wave. This results in a power-law temporal decay and a non-thermal spectrum widely accepted to be caused by synchrotron emission. %
Since the launch of the \emph{Swift} satellite \cite{gehrels2004:ApJ611}, it has become clear that this model for GRBs cannot, in its current form, explain the complexity of observed light curves --  \emph{Swift}'s fast-slew capability allows for much earlier observations and a more elaborate picture of the evolution of the emission, particularly in the X-ray regime using the X-ray Telescope (XRT, \citeNP{burrows2005:SSRv120}). 
The unexpected features detected, such as steep decays, plateau phases (e.g., \citeNP{tagliaferri2005:Natur436,nousek2006:ApJ642,obrien2006:ApJ647}) and a large number of X-ray flares (e.g., \citeNP{burrows2007:RSPT365,chincarini2007:ApJ671,falcone2006:ApJ641,falcone2007:ApJ671}) have revealed the complexity of these sources up to $\sim$1 day since the initial event, which is yet to be fully understood.

Prior to \emph{Swift}, X-ray coverage typically began at $\sim$ 0.5-1.5\,days after the GRB event, and X-ray flares were detected in only a few bursts (e.g., GRB\,970508, \citeNP{piro1998:A&A331L}; GRB\,011121 \& GRB\,011211, \citeNP{piro2005:ApJ623}). However, they are now observed in $\sim 50$\% of all \emph{Swift} GRBs, typically superposed on the early light curve steep decay and plateau phases. A clear clustering exists in the total flare distribution in time, with the vast majority occurring at early times up to $\sim$1,000\,seconds \cite{chincarini2007:ApJ671,burrows2007:RSPT365}. 
Various possible explanations have been put forward including external origins due to patchy shells \cite{meszaros1998:apj499,kumar2000:ApJ535}, refreshed shocks (e.g., \citeNP{rees1998:ApJ496L,zhang2002:ApJ566}), density fluctuations \cite{wang2000:ApJ535,dai2002:ApJ565L} or continued central engine activity \cite{dai1998:A&A333L,zhang2002:ApJ566}, though a consensus has been slow to emerge.
A very small number of GRBs have exhibited flares on much later timescales of $10^4$ to $10^5$ seconds, or approximately one day after the prompt event (including, in the optical, pre-\emph{Swift} GRB\,000301C, \citeNP{sagar2000:BASI28}).
These late flares, like the early flares, are difficult to accommodate within the external shock model if the width, $\Delta t$, is smaller than the observing timescale, $t$, as is often the case \cite{lazzati2002:A&A396,lazzati2007:MNRAS375}. They can also be difficult to accommodate within the internal shock model because that would require prolonged activity out to late times or a restarting of the central engine, though a number of methods for doing so have been suggested \cite{zhang2006:ApJ642}.

Here we examine the X-ray light curves of \emph{Swift} GRBs for such late flares and compare their properties with those of early flares to investigate whether they are simply the tail of the early flare distribution or form a different sample. In section \ref{section:sample} we introduce our sample while in section \ref{section:analysis} we discuss the methods and results of our temporal and spectral analyses. In section \ref{section:discussion} we discuss these results in the overall context of the blast wave model of GRBs and the internal/external flare models. We summarise our findings in section \ref{section:conclusion}.
Throughout, we use the convention that a power-law flux is given as $F \propto t^{-\alpha} \nu^{-\beta}$ where $\alpha$ is the temporal decay index and $\beta$ is the spectral index. All uncertainties are quoted at the $1\sigma$ confidence level.


\section{Sample selection}\label{section:sample}

\begin{figure*}[!]
 \centering 
\begin{minipage}{67mm}
\resizebox{67mm}{!}{\includegraphics[angle=-90]{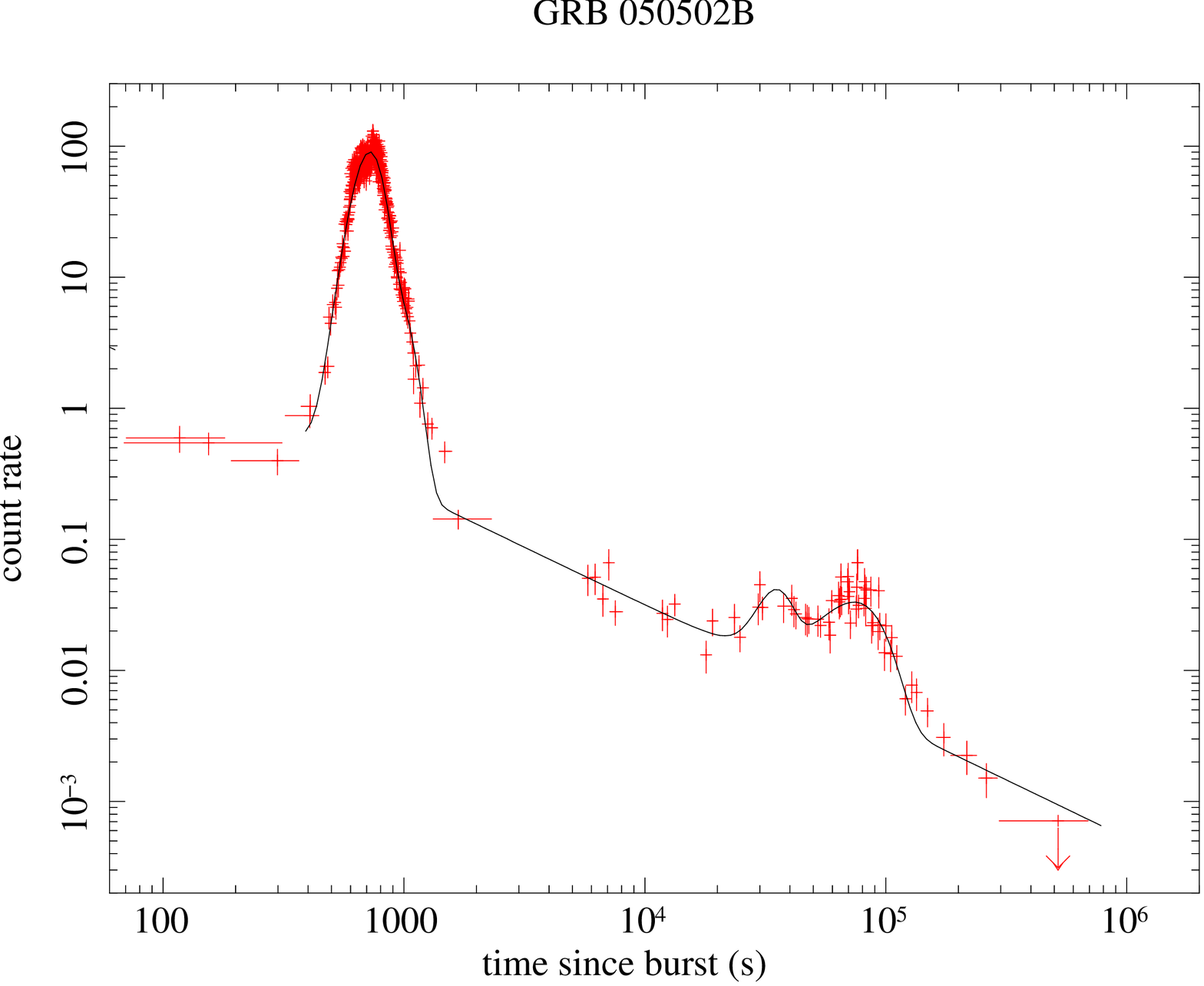} }
\end{minipage}
\hspace{3mm}
\vspace{3mm}
\begin{minipage}{67mm}
\resizebox{67mm}{!}{\includegraphics[angle=-90]{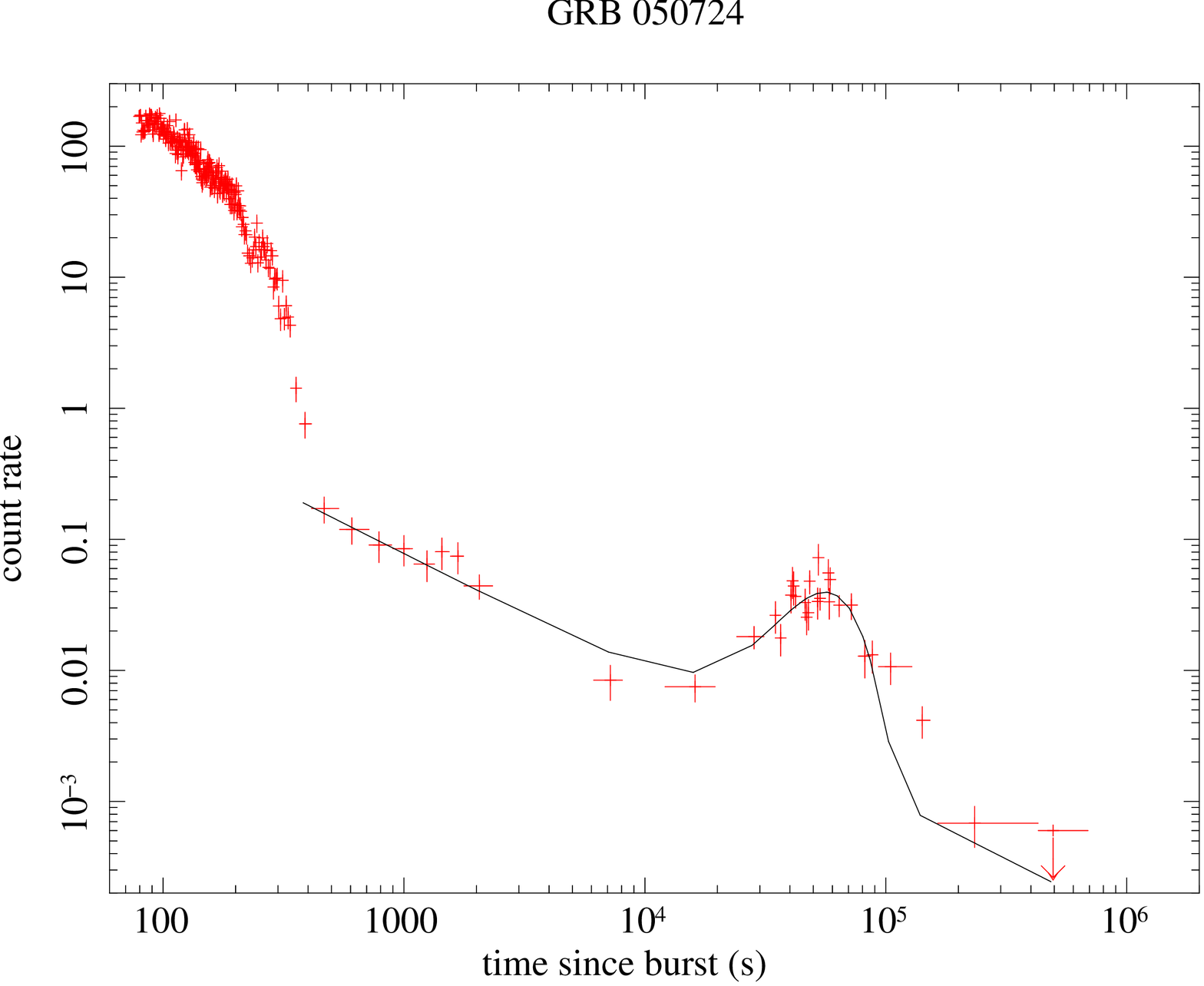} }
\end{minipage}
\vspace{3mm}
\begin{minipage}{67mm}
\resizebox{67mm}{!}{\includegraphics[angle=-90]{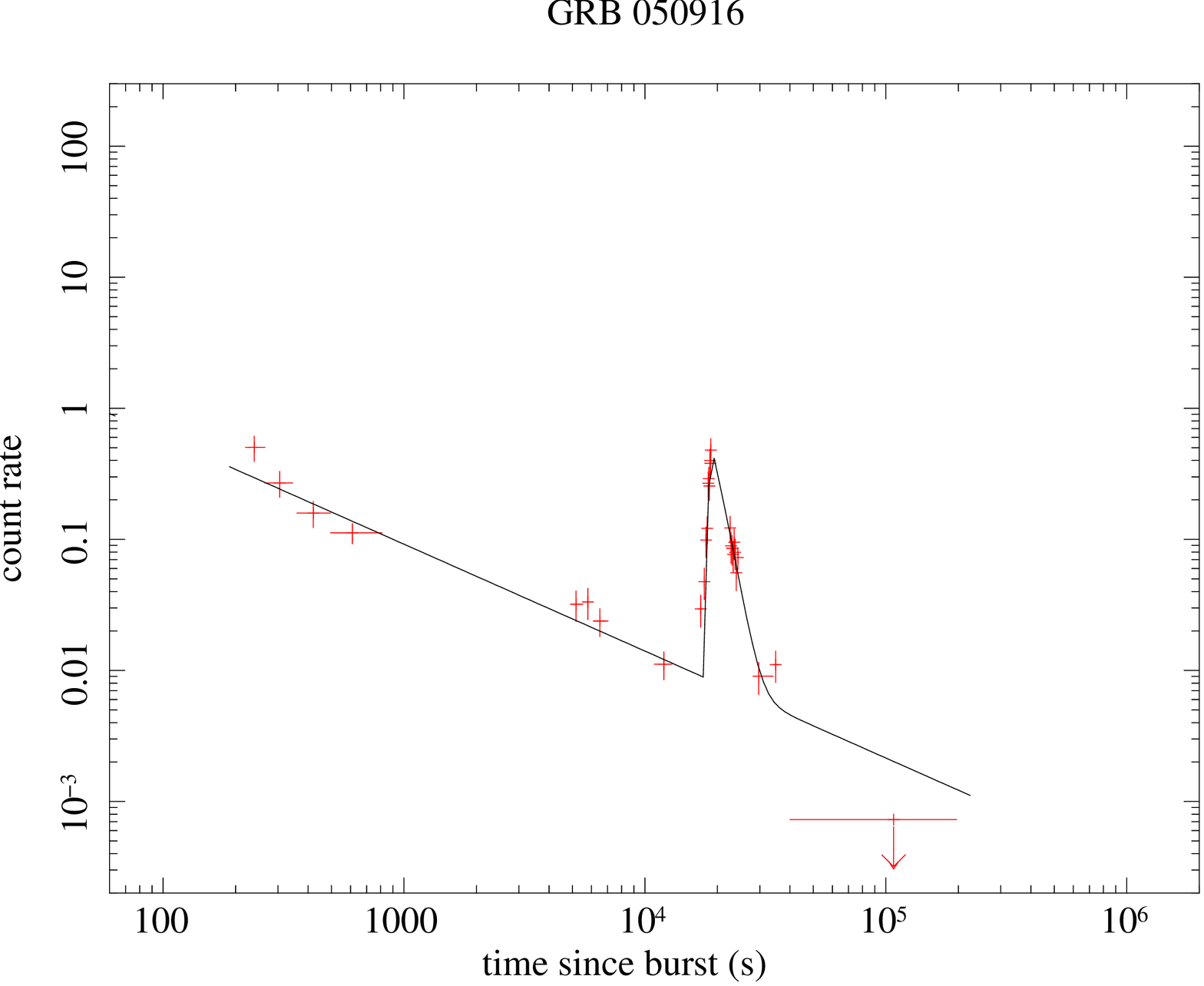} }
\end{minipage}
\hspace{3mm}
\begin{minipage}{67mm}
\resizebox{67mm}{!}{\includegraphics[angle=-90]{070107.ps} }
\end{minipage}
\vspace{3mm}
\begin{minipage}{67mm}
\resizebox{67mm}{!}{\includegraphics[angle=-90]{070311.ps} }
\end{minipage}
\hspace{3mm}
\begin{minipage}{67mm}
\resizebox{67mm}{!}{\includegraphics[angle=-90]{070318.ps} }
\end{minipage}
\begin{minipage}{67mm}
\resizebox{67mm}{!}{\includegraphics[angle=-90]{070429a.ps} }
\end{minipage}
\hspace{3mm}
\begin{minipage}{67mm}
\hspace{30mm}
\end{minipage}
\caption{Power-law plus Gaussian fits to the XRT light curves of each of the bursts in our sample, except for GRB\,050916 which is better fit by a power-law plus FRED like flare.} 
\label{lightcurves} 
\end{figure*}

From a visual inspection of the pre-reduced \emph{Swift} XRT light curves in the on-line repository \cite{evans2007:A&A469} up to the end of December 2007, we identified a sample of 7 bursts which clearly exhibit very late ($\gtrsim 1 \times 10^4$\,s) flares (Figure \ref{lightcurves}). The definition of the cutoff at $1 \times 10^4$\,s is entirely arbitrary but chosen so as to study a sample of the latest flares possible. As the redshifts of the majority of our sample are unknown, we are unable to correct the time of the flare to the rest frame time; we therefore caution that these flares are not necessarily at such late times intrinsically. However, even if we assume that these bursts are at the average \emph{Swift} redshift of $\sim 2.8$ \cite{jakobsson2006:A&A447}, they are still on the tail of the temporal distribution \cite{chincarini2007:ApJ671}. These bursts are sampled well enough to allow an unambiguous identification of a flare, i.e., sparsely sampled bursts where the data could not rule out a flare were not used except for the case of GRB\,070311 where the presence of the late flare is confirmed via optical observations. 

We only include bursts where we can unambiguously obtain the underlying temporal decay so that we can better constrain the flare parameters. This approach favours more obvious, sharper, stronger flares, while neglecting slower, dimmer flares. Our selection also gives a biased sample towards bursts more easily followed up with \emph{Swift}, which are bright to late times such that they can be observed out to $1 \times 10^5$\,s and a flare detected. This also favours bursts of shallower temporal decay as they are more likely to be observed out to these late times.
Despite these biases, the discovery of flares due to internal processes at late times would place strong constraints on models.

The sample is summarised as follows: 
GRB\,050502B, as previously discussed by \citeANP{falcone2006:ApJ641}, displays an energetic early flare as well as two overlapping late flares.  
GRB\,050724 displays a late flare at X-ray and optical wavelengths \cite[respectively]{campana2006:A&A454,malesani2007:A&A473}. We note also that this is considered a short burst with T$_{90} = 3$\,s over 15-350\,keV \cite{krimm2005:GCN3667} and 2\,s in softer energy bands \cite{remillard2005:GCN3677}. 
GRB\,050916 displays an obvious, sharp late time flare, though the temporal coverage is poor.
GRB\,070107 displays two early flares as well as the late flare \cite{stamatikos2007:GCNR25}. 
GRB\,070311 was initially detected by INTEGRAL so only has XRT coverage past 7000\,s, however, coverage at optical wavelengths has confirmed the suspected flare and allowed a full study of this burst \cite{guidorzi2007:GCNR41,guidorzi2007:A&A474}. 
GRB\,070318 displays both an early and late flare which was also observed by UVOT in the optical \cite{cummings2007:GCNR40} while GRB\,070429A has a single late flare \cite{cannizzo2007:GCNR52}.

\begin{table}[ht]
  \begin{center} 	
    \caption{The spectroscopic redshifts, $z$, Galactic absorption, $N_{\mathrm{H}}$(Galactic) \protect\cite{kalberla2005:A&A440} and total fitted absorption, $N_{\mathrm{H}}$(Total), for the bursts in our sample.} 	
    \label{sample-info} 	
    \begin{tabular}{l l l l l } 
      \hline \hline
      GRB     & $z$     & $N_{\mathrm{H}}$(Galactic)     & $N_{\mathrm{H}}$(Total)    \\
              &         & $\times 10^{22}$\,cm$^{-2}$ & $\times 10^{22}$\,cm$^{-2}$       \\
      \hline
      050502B &   --    & 0.0359            &  $\leq$ 0.11   \\
      050724  & 0.258$^{1}$ & 0.140  &   0.75$^{+0.5}_{-0.2}$ \\
      050916  &   --    & 0.810              &   1.3$^{+0.5}_{-0.3}$ \\
      070107  &   --    & 0.299             &   0.47 $\pm$ 0.03\\
      070311  &   --    & 0.236             &   0.6 $\pm$ 0.1  \\
      070318  & 0.84$^{2}$  & 0.0144 &  0.33 $\pm$ 0.03  \\
      070429A &   --    & 0.0839            &  0.22 $\pm$ 0.06  \\
      \hline
    \end{tabular}
\begin{list}{}{}
\item[]$^{1}$ \citeN{prochaska2005:GCN3700}. $^{2}$ \citeN{chen2007:GCN6217}.
\end{list}
  \end{center}
\end{table}


\section{Analysis \& results}\label{section:analysis}

\subsection{Spectra}

The XRT event data for these bursts were initially processed with the FTOOL, \texttt{xrtpipeline (v0.11.4)}. Source and background spectra from the Photon Counting mode data (PC; \citeNP{hill2004:SPIE5165}) were extracted and bad columns corrected for, where necessary. Pile-up was tested for but, due to the lateness of the data, was not an issue. Spectra were binned to have $\geq$20 photons per bin and the \texttt{v010} response matrices were used. The spectra, from 0.3-10.0\,keV, were fit with absorbed power-laws in \texttt{Xspec 11.3.2} using $\chi^{2}$ statistics. The Galactic value of Column Density, $N_{\mathrm{H}}$, was taken from \citeN{kalberla2005:A&A440}. 

Due to the lateness and hence dimness of these flares, spectral analysis of the individual flares similar to that done by \citeN{falcone2007:ApJ671} was not possible. Instead, we fit the spectra of the underlying afterglow so as to determine the electron energy distribution indices, $p$,  via spectral indices, $\beta$. As the relationships between $p$ and  $\beta$ \cite{zhang2004:IJMPA19} hold only for the underlying afterglow and not the flaring region, the spectra extracted for analysis were taken from the start of the temporal power-law decay (i.e., after early flares or steep decay phases) and the late flares were eliminated. This leads, in some cases, to spectra with low total counts and hence poorly constrained fit parameters as can be seen from the results of the fits (Tables \ref{sample-info} \& \ref{indices}).

\subsection{Light curve modelling}

Our light curve analyses are carried out on the pre-reduced, XRT light curves from the on-line repository \cite{evans2007:A&A469}.We model the light curves with a combination of power-law decay (Table \ref{indices}) and Gaussian flares with peak, $t$, width, $\sigma$, and full width at half maximum, $\Delta t = 2 \sqrt{2 \ln2} \sigma \sim 2.3548 \sigma$ (Table \ref{flares}). These fits (Figure \ref{lightcurves}) are used to find the relative temporal and flux variability ($\Delta t / t$ and $\Delta F / F$ where $\Delta F$ is the excess flux of the flare over that of the underlying power-law).

In one burst, GRB\,070107, a single power-law fit was unsatisfactory so a smoothly broken power-law was adopted. The break at ($1.7 \pm 0.4$) $\times 10^5$\,s, after which the temporal index drops to $\alpha = 1.9 \pm 0.2$, does not affect our analysis as it occurs after the flare. 
In the case of GRB\,050916 a FRED (fast rise exponential decay) -like flare was a better fit than a Gaussian, returning a temporal power-law index consistent with that of the Gaussian fit. This index overestimates the final data point, a $3\sigma$ upper limit, but we can not imply a break from this as it is likely just an outlier. The FRED-like profile of the flare had very fast rise ($\sim 1800$\,s) and decay ($\sim 2100$\,s) times peaking at $\sim 1.9 \times 10^4$\,s. The analysis however, was carried out on the parameters of the Gaussian based fit to be comparable with the rest of the sample. 
GRB\,070429A, after an initial steep decay phase, has little data until the flare so the underlying power-law is more uncertain than the errors would suggest as we are unable to rule out other flares affecting the value. The fit also overestimates the final data point, which we did not include in the fit as it introduced a solution in the fit of the flare, not believed to give the true values of the flare parameters. The overestimate may be related to a break which could not be confirmed from the limited data available.

\begin{table}[ht]	
  \begin{center} 	
    \caption{Indices of the underlying temporal power-law, $\alpha$, the spectral power-law, $\beta$, the electron energy distribution, $p$ and the value of the temporal decay it predicts, $\alpha_{p}$.} 	
    \label{indices} 	
    \begin{tabular}{l c c c c} 
      \hline \hline
      GRB     & $\alpha$        & $\beta$           & $p$              & $\alpha_{p}$       \\
      \hline
      050502B & 0.89 $\pm$ 0.04 & 1.0 $\pm$ 0.2     & 2.0 $\pm$ 0.4    & 1.0 $\pm$ 0.2    \\
      050724  & 0.93 $\pm$ 0.09 & 1.4 $\pm$ 0.5     & 2.8 $\pm$ 1.0    & 1.6 $\pm$ 0.6   \\
      050916  & 0.71 $\pm$ 0.05 & 0.8 $\pm$ 0.2     & 1.6 $\pm$ 0.4    & 0.9 $\pm$ 0.2   \\
      070107  & 1.03 $\pm$ 0.03 & 1.27 $\pm$ 0.07   & 2.54 $\pm$ 0.14  & 1.41 $\pm$ 0.08   \\
      070311  & 1.2 $\pm$ 0.5   & 1.2 $\pm$ 0.2     & 2.4 $\pm$ 0.4    & 1.3 $\pm$ 0.2  \\
      070318  & 1.10 $\pm$ 0.15 & 1.4 $\pm$ 0.1     & 2.8 $\pm$ 0.2    & 1.6 $\pm$ 0.1   \\
      070429A & 0.38 $\pm$ 0.05 & 1.10 $\pm$ 0.15    & 2.2 $\pm$ 0.3    & 1.15 $\pm$ 0.15  \\
      \hline
    \end{tabular}
  \end{center}
\end{table}

\begin{table}[ht]	
  \begin{center} 	
    \caption{Peak time, $t$, Gaussian width $\sigma$, relative temporal and flux variability, $\Delta t /t$ and $\Delta F/ F$, of the late flares.} 	
    \label{flares} 	
    \begin{tabular}{l c c c c } 
      \hline \hline
      GRB      & $t$             & $\sigma$       & $\Delta t /t$    & $\Delta F/ F$    \\
               & $\times 10^4$\,s & $\times 10^4$\,s        &  &     \\
      \hline 
      050502B & 3.52 $\pm$ 0.14 & 0.61 $\pm$ 0.15   & 0.36 $\pm$ 0.09  & 3.0$^{+1.2}_{-1.9}$  \\
              & 7.53 $\pm$ 0.18 & 2.2  $\pm$ 0.2    & 0.69 $\pm$ 0.06  & 5.3$^{+1.3}_{-3.1}$   \\
      050724  & 5.72 $\pm$ 0.16 & 1.89 $\pm$ 0.16   & 0.78 $\pm$ 0.07  & 21$^{+8}_{-16}$  \\
      050916  & 2.040 $\pm$ 0.005 & 0.12 $\pm$ 0.01 & 0.14 $\pm$ 0.01  & 70 $\pm$ 40  \\  
      070107  & 8.9 $\pm$ 0.2   & 1.3 $\pm$ 0.2     & 0.35  $\pm$ 0.05 & 0.9$^{+0.3}_{-0.4}$  \\
      070311 &  15.8 $\pm$ 1.7   & 7.0$^{+2.0}_{-1.2}$  & 1.0$^{+0.3}_{-0.2}$   & $\leq$8.1   \\
      070318 & 17$^{+4}_{-13}$    & 16$^{+9}_{-5}$       & 2.2$^{+2.1}_{-0.9}$   & 1.1 $\pm$ 0.2  \\
      070429A & 26.27 $\pm$ 0.05 & 1.4$^{+0.8}_{-0.3}$  & 0.12$^{+0.07}_{-0.03}$  & 1.9$^{+1.2}_{-1.5}$   \\
      \hline
    \end{tabular}
  \end{center}
\end{table}


\section{Discussion}\label{section:discussion}

The early behaviour of the bursts is varied; of the 5 sources with early observations, 2 exhibit a steep decay phase while the other 3 display early flares over a power law decay. Since the redshifts of the majority of our sample are unknown, it is impossible to confirm the apparent agreement of the time of the flares at $t_{\rm{avg}} \sim 1 \times 10^{5}$\,s, though this is certainly affected by our definition of late as $\gtrsim 1 \times 10^4$\,s.

\subsection{Spectral and temporal indices and the blast wave model}\label{dis:alphas}

The underlying temporal indices of the bursts range from $\alpha \sim 0.4$ to $\alpha \sim 1.2$, which are quite shallow,  suggesting that these flares might occur during plateau phases (e.g., \citeNP{nousek2006:ApJ642,obrien2006:ApJ647}). To test this we compare the observed values of temporal indices with those derived from the spectral indices, assuming the standard closure relations \cite{zhang2004:IJMPA19}.
We used these relations to calculate the electron energy distribution index, $p$, and the predicted values of the temporal slope, $\alpha_{p}$, from the measured spectral indices of the X-ray spectrum, $\beta$ (Table \ref{indices}). We have assumed that the cooling and peak frequencies are below the X-rays, $\nu_{\mathrm{c,m}} < \nu_{\mathrm{X}}$, to estimate the shallowest slopes possible for a given spectral index (steeper slopes may be estimated if $\nu_{\mathrm{m}} < \nu_{\mathrm{X}} < \nu_{\mathrm{c}}$).

In all cases the predicted value of temporal decay overestimates the measured value, though in 4 of the 7 cases the values overlap within the  $1\sigma$ level, and 1 at the $2\sigma$ level. It should however be cautioned that many of the predicted temporal decays have significant errors because of the low total counts of their X-ray spectra. In the case of GRB\,070429A, there is no overlap until $3.7\sigma$, implying that it does not correspond to the regular decay in the blast wave model. It can be explained by energy injection \cite{nousek2006:ApJ642} of the form $E \propto t^{\sim 0.6}$, which however, would cause a break to regular behaviour which is not observed. Given the poor temporal sampling of this burst and the uncertainty of the underlying power-law it is difficult to make conclusive statements.

In the case of GRB\,070107 the overlap is similarly marginal, at the $3.5\sigma$ level. This is the one burst in our sample where we are able to say there is a break at late times ($\sim 1.7 \times 10^5$\,s). This may be a jet break, though the observed slope of $\alpha = 1.9 \pm 0.2$ is shallower than the expected  $\alpha_{p} = p = 2.54 \pm 0.14$ but may be rolling over to the asymptotic value expected. Another explanation may be that X-ray frequency is between that of the peak and cooling frequencies, $\nu_{\mathrm{m}} < \nu_{\mathrm{X}} < \nu_{\mathrm{c}}$, in a constant density circumburst medium: in this case we find $p = 3.54 \pm 0.12$ implying $\alpha_{p} = 1.91 \pm 0.06$ which agrees well with our post break slope. The break could then be interpreted as being due to the cessation of continued energy injection of the form $E \propto t^{\sim 0.6}$.

\begin{figure} 
  \centering 
  \resizebox{\hsize}{!}{\includegraphics[angle=-90]{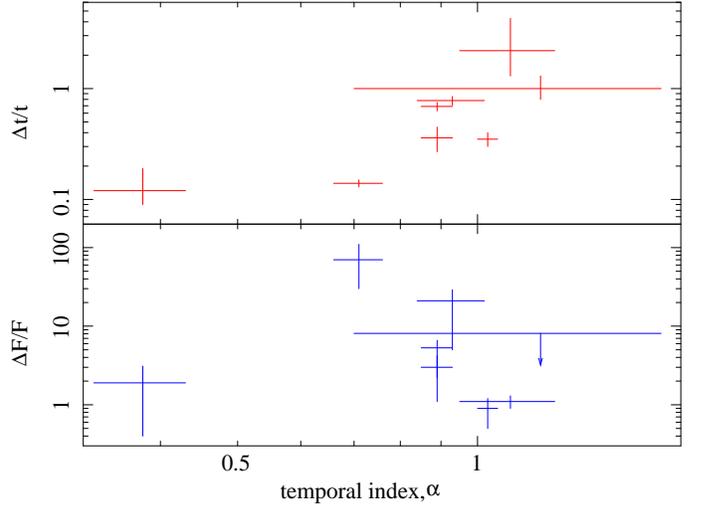} }
  \caption{Relative temporal and flux variabilities, $\Delta t /t$ (upper) and $\Delta F/ F$ (lower;  Table \ref{flares}) versus the underlying temporal power-law decay indices (Table \ref{indices}) show no correlation.} 
  \label{alpha_var} 
\end{figure}

The $1\sigma$ agreement of the majority of the bursts does not preclude the possibility of energy injection in these cases, especially given the significant errors. It is most likely that the bursts are taken from both samples. The properties of the flares themselves, i.e., relative flux variability, $\Delta F /F$, and temporal variability, $\Delta t /t$, do not seem to be dependent on the underlying temporal decay power-law (Figure \ref{alpha_var}), though with such a small sample this is inconclusive.

\subsection{The origin of late flares: internal vs. external shocks}

Applying kinematic arguments, \citeN{ioka2005:ApJ631} place limits on the timescale and flux amplitude variabilities ($\Delta t /t$ and $\Delta F/ F$)  allowed by various flare, or bump, afterglow origin models (i.e., external shocks): patchy shells, refreshed shocks, on-axis density fluctuations and off-axis density fluctuations of many regions. 
The limits they find are $\Delta t /t \ge 1$, $\Delta t /t \ge 0.25$, $\Delta F/ F \le 1.6 \Delta t /t$ and $\Delta F/ F \le 24 \Delta t /t$ respectively, assuming $F/\nu F_{\nu} \sim 1$ and the fraction of cooling energy, $f_{c} \sim (\nu_{\mathrm{m}}/\nu_{\mathrm{c}})^{(p-2)/2} \sim 1/2$ (\citeANP{ioka2005:ApJ631} section 3.2, section 3.3, equation 7, equation A2).
\citeN{chincarini2007:ApJ671} plot their sample's properties on these regions and find that while all flares may be explained as being due to internal shocks ($\Delta t /t < 1$) caused by long-lasting central engine activity, about half could be due to refreshed shocks of external shock origin and $\sim 15$\,percent could only be explained by prolonged central engine activity.

\begin{figure} 
  \centering 
  \resizebox{\hsize}{!}{\includegraphics[angle=-90]{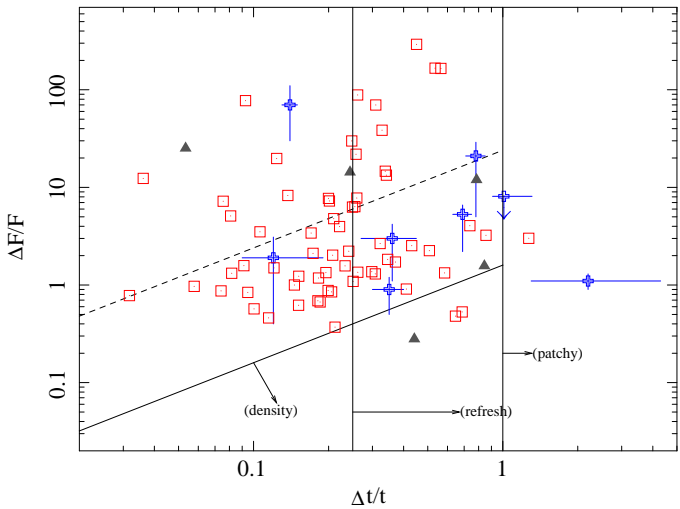} }
  \caption{Our relative temporal and flux variabilities, $\Delta t /t$ and $\Delta F/ F$ (Table \ref{flares}), of the late flares (blue error bars) plotted on the allowable kinematic regions \protect\cite{ioka2005:ApJ631}. Also plotted are the early (squares) and late (triangles) sample of \protect\citeN{chincarini2007:ApJ671}. The lines shown are for bumps due to patchy shells ($\Delta t /t \ge 1$), refreshed shocks ($\Delta t /t \ge 0.25$), on-axis density fluctuations ($\Delta F/ F \le 1.6 \Delta t /t$) and off-axis density fluctuations of many regions (dashed line; $\Delta F/ F \le 24 \Delta t /t$).} 
  \label{ioka} 
\end{figure}

We have compared the values for $\Delta t /t$  and  $\Delta F/ F$ of the late flares with those published by \citeANP{chincarini2007:ApJ671} and with the theoretical limits of \citeANP{ioka2005:ApJ631}. We find that these properties of the late flares in our sample are in agreement with the distribution of values found for early and late flares by  \citeANP{chincarini2007:ApJ671} (Figure \ref{ioka}). 
In this figure, the differences between the values for late flares in the \citeANP{chincarini2007:ApJ671} sample (triangles) and ours (blue error bars), show that there are some differences between the two analyses.  While the properties of GRB\,050724 agree within our errors, those of GRB\,050502B and GRB\,050916 appear to be the most deviant.
In the case of GRB\,050502B, \citeANP{chincarini2007:ApJ671} fit the underlying temporal power-law as a broken one, while our light curve does not support this so is fit by a single power-law. We also find that the early flare is better fit by two Gaussians as opposed to one. Both of these differences cause an offset of the properties in question. 
For GRB\,050916 we find that the late flare is well fit by only one Gaussian as opposed to the two used by \citeANP{chincarini2007:ApJ671}.

We find that all the late flares in our sample, bar that of GRB\,070318 with $\Delta t /t > 1$ and which is also observed in the optical \cite{cummings2007:GCNR40}, may be explained by internal shocks, though only the FRED-like flare of GRB\,050916 can be explained only by internal shocks. GRB\,050916 lies to the top left of the distribution in Figure \ref{ioka} and while it remains within the early-flare parameter space, it shows more extreme properties than a typical early flare. This can be seen in the shape of its flare, which has a very rapid rise-time and a FRED-like shape rather than a Gaussian. Internal shocks are almost certainly responsible for this particular flare as density enhancements or refreshed shocks would not be able to create such a fast rise. The late flares of both GRB\,050724 and GRB\,070311 have also been observed in the optical \cite[respectively]{malesani2007:A&A473,guidorzi2007:A&A474} and this apparent achromatic nature suggests external origins. This is certainly consistent with their flare properties as shown in Figure \ref{ioka}. Though, these flares do lie in the expected range for an internal origin and are not distinct from those of early-time flares so this is inconclusive.

The majority of late flares in our sample, similar to the early flares, may be explained both by internal shocks or one of the external models. The evidence from this limited sample suggests that these flares are no different from the sample of early flares. They are, most likely, late examples on the tail of the distribution of early flares, though this needs to be confirmed by a larger sample with known redshifts as it becomes available.


\section{Conclusions}\label{section:conclusion}

We have examined the \emph{Swift}-XRT light curves of GRBs up to December 2007, and identified a sample of 7 bursts which clearly exhibit late time flares. The early behaviour of these bursts, where observed, is varied: either steep decay, or, flaring overlaid on a power-law decay. Some of the underlying power-law decays, at the time of the late flares, are probably due to continued energy injection, while the remainder are probably the normal decay phase of the afterglow. As the flares occur at late times and may obscure possible underlying breaks (energy injection, jet or spectral) we can only confirm a break in one burst.

The evidence does not suggest that these flares are any different from the sample of early flares, and hence they are most likely late examples on the tail of the distribution. Like the sample of early flares, most can be explained both by internal shocks or external models. We should caution however, that the sample presented here is not unbiased and there are certainly selection effects which favour brighter bursts, more easily followed up with \emph{Swift}, and more obvious, sharper, stronger flares.
If this distribution is confirmed, as the sample with known redshifts and well sampled light curves grows, it would imply that internal processes produce significant flares up to a day after the prompt event. Hence, in some cases at least, prolonged activity out to late times or a restarting of the central engine is required.


\begin{acknowledgements}
We thank R. Willingale and P. Evans for useful discussions on the XRT light curves. 
We thank the referee for their comments.
PAC and RAMJW gratefully acknowledge the support of NWO under grant 639.043.302.
PAC acknowledges the support of the University of Leicester SPARTAN exchange visit programme, funded by the European Union Framework 6 Marie Curie Actions. 
RLCS and OG acknowledge support from STFC.
AJvdH is supported by an appointment to the NASA Postdoctoral Program at the NSSTC, 
administered by Oak Ridge Associated Universities through a contract with NASA.
This work made use of data supplied by the UK Swift Science Data Centre at the University of Leicester funded by STFC and through the High Energy Astrophysics Science Archive Research Center Online Service, provided by the NASA/Goddard Space Flight Center.  
\end{acknowledgements}

\end{document}